# Enhancing Performance of Subject-Specific Models via Subject-Independent Information for SSVEP-Based BCIs

Mohammad Hadi Mehdizavareh[1], Sobhan Hemati[1], Hamid Soltanian-Zadeh[1,2]

[1]CIPCE, School of Electrical and Computer Engineering, College of Engineering, University of Tehran, Tehran, Iran

[2]Medical Image Analysis Laboratory, Departments of Radiology and Research Administration, Henry Ford Health System, Detroit, MI 48202, USA

## Abstract

Recently, steady-state visual evoked potential (SSVEP)-based brain-computer interface (BCI) has attracted much attention due to its high information transfer rate (ITR) and increasing number of targets. However, the performance of SSVEP-based methods in terms of accuracy and time length required for target detection can be improved. We propose a new canonical correlation analysis (CCA)-based method to integrate subject-specific models and subject-independent information and enhance BCI performance. To optimize hyperparameters for CCA-based model of a specific subject, we propose to use training data of other subjects. An ensemble version of the proposed method is also developed and used for a fair comparison with ensemble task-related component analysis (TRCA). A publicly available 35-subject SSVEP benchmark dataset is used to evaluate different methods. The proposed method is compared with TRCA and extended CCA methods as reference methods. The performance of the methods is evaluated using classification accuracy and ITR. Offline analysis results show that the proposed method reaches highest ITR compared with TRCA and extended CCA. Also, the proposed method significantly improves performance of extended CCA in all conditions and TRCA for time windows greater than 0.3 s. In addition, the proposed method outperforms TRCA for low number of training blocks and electrodes. This study illustrates that adding subject-independent information to subject-specific models can improve the performance of SSVEP-based BCIs.

# 1. Introduction

Brain-computer interface (BCI) systems have been recognized as a new Communication channel for humans, especially severely disabled individuals [1-3]. One of the most important applications of BCI is character speller system which allows disabled individuals to communicate with their surrounding environment [2]. Electroencephalography (EEG) is a noninvasive, low cost, and simple modality, widely used to implement BCI spellers [4]. In recent years, steady-state visual evoked potential (SSVEP)-based BCI spellers have attracted much more attention compared with other BCI systems including motor imagery and P300. This is because of their high information transfer rate (ITR), less user training, and the ability to deal with problems with a large number of classes [4-7].

There are many target coding methods in SSVEP-based BCIs, among which frequency coding is a popular method to encode targets [8, 9]. Several methods have been proposed to combine phase and frequency coding approaches [10-12]. The most discriminative method is joint frequency-phase modulation (JFPM) method which assigns different frequencies and phases to two adjacent targets [12]. Target identification is another crucial issue in SSVEP-based BCIs, for which numerous methods have been proposed. Initially, single-channel methods were presented based on power spectral density analysis (PDSA) [13-14] and then multiple channel methods were introduced to improve the signal to noise ratio (SNR) of SSVEP response. In these methods, channels are combined using appropriate spatial filters so that common noises in the channels are reduced and the quality of SSVEP response is improved. Some powerful examples of such methods are minimum energy combination (MEC) [15], Maximum contrast combination (MCC) [15], and canonical correlation analysis (CCA) [16]. Although these methods are widely used because of their simplicity and free training attribute, they only detect frequency. They are unable to discriminate two different phases [11], and their performance degrades in short time windows due to the presence of the background noise in the EEG signal. To solve these problems, incorporating individual calibration data has been proposed [12, 17-20]. Extended CCA method was introduced to combine CCA coefficient with the Pearson correlation coefficients among the test and training data [12]. Multiway CCA (MwayCCA) [17], L1-regularized MwayCCA [18], and multiset CCA (MsetCCA) [19] were proposed to optimize artificial sine-cosine reference signals embedded in CCA using training trials of each subject. Also, task-related component analysis (TRCA) was suggested to enhance the SNR of SSVEP response using optimized spatial filters [20]. TRCA extracts task-related components by maximizing the reproducibility during the task period [21]. Comparison studies have shown that extended CCA and TRCA methods provide the best performance in terms of classification accuracy and ITR, especially in short time windows [20, 22]. Thus, these two methods are selected as reference methods to be compared with the proposed method in this paper.

From training point of view, target identification methods can be classified to three main categories [23]: 1- training free methods such as PSDA and CCA, which do not need any calibration data; 2- subject-specific training methods such as extended CCA and TRCA, in which calibration data are collected for each subject and the parameters of the algorithm are optimized individually; 3- subject-independent training methods like transfer template-based CCA (tt-CCA) [24], which use the training data of existing subjects to create a fixed model for a new subject.

In this paper, a new CCA-based method is proposed which exploits both subject-specific and subject-independent training methods to enhance performance of a BCI system. A publicly available 35-subject SSVEP benchmark dataset [25] is used to evaluate the proposed method. First, the most informative CCA-based correlation coefficients are found using a subject-independent training method and then the selected coefficients are used for a new subject. Also, an ensemble version of the CCA-based method is introduced in which a linear combination of correlation coefficients derived from basic spatial filters and ensemble spatial filters are used to construct the final feature for target identification.

The remainder of the paper is organized as follows. Section 2 introduces benchmark dataset and data preprocessing applied to all methods and reviews standard CCA, extended CCA, and TRCA methods. Then, the basic and ensemble version of the proposed algorithm is described in details, and finally, filter bank analysis is provided. Section 3 presents the experimental results. In section 4, the difference between the proposed algorithm and extended CCA method is discussed, and the advantages of our method over other methods are shown. Section 5 concludes the paper.

## 2. Methods

### 2.1. Benchmark dataset

In this study, benchmark dataset introduced in [25] has been used. This dataset is freely available for BCI community to simplify the comparison among different SSVEP response detection algorithms. This dataset has been collected from 35 subjects (17 females, 18 males, mean age 22 years, 27 naïve, and 8 experienced). The designed experiment includes a 40-target speller system which uses JFPM method to encode characters with 0.2 Hz frequency difference and $0.5\pi$ phase difference between two neighboring targets. Also, the frequency interval used in this task is in the range of [8, 15.8 Hz]. It has been shown that phase interval of $0.35\pi$ leads to the best performance of the BCI system [12], so the method proposed in [12, 25] is used in this study to shift the EEG data circularly such that the phase difference is converted to $0.35\pi$. For each subject, the task consists of six blocks, and each block includes 40 trials (one trial for each target) which are randomly presented through the LCD to the subjects. In each trial, a visual cue (red square) is shown on the screen for 0.5 s and the subjects are asked to follow the cue target on the screen using their eyes. As the cue disappears, all 40 targets start flickering simultaneously for 5 s and when the stimuli is finished the screen is blank for 0.5 s before the next trial starts and therefore each trial lasts 6 s. In every block, the subjects are asked to avoid blinking during stimulus presentation. To avoid eye fatigue, there is a rest for several minutes between two successive blocks.

The EEG data were acquired from 64 channels using Synamps2 system (Neuroscan Company) with 1000 Hz frequency sampling. The electrodes were located according to the international 10-20 system. The ground electrode was located somewhere between Fz and FPz electrodes, and the reference electrode was placed at vertex. The amplifier frequency passband was between 0.15 and 200 Hz, and the electrode impedances were kept less than 10 kΩ. Also, during data recording a notch filter was used to remove the 50 Hz power line noise. The synchronous signal generated by the stimulus program was sent to the amplifier and recorded on an event channel synchronized to the visual cue onset. To reduce the data size, all EEG epochs were down-sampled to 250 Hz. Further details about the dataset can be found in [25].

### 2.2. Data Preprocessing

The first step for EEG data preprocessing is channel selection. The SSVEP topographic scalp maps show high activity over parietal and visual areas [26, 27], so based on previous studies [12, 25], nine electrodes located in these areas (O1, O2, Oz, PO3, PO4, PO5, PO6, POz, and Pz) are selected. By taking into account 140 ms latency delay in the visual system [12, 28], for a time window with length Tw s, all epochs are extracted in the interval [0.14 s  0.14+Tw s] in which time 0 indicates stimulus onset. Then, all segmented epochs are band-pass filtered from 6 Hz to 90 Hz using a zero-phase Chebyshev Type II Infinite impulse response (IIR) filter. The filtfilt() function in MATLAB is used to implement zero-phase forward and reverse filtering.

## 2.3. Reference Methods

### 2.3.1. Standard CCA Method

CCA is a statistical multivariate method to maximize the correlation between two sets of variables and has been widely used in SSVEP-based BCIs for frequency detection [16, 29]. Let $f_K$, $F_s$, $N_t$, $M$, $K$, and $N_h$ denote k-th stimulus frequency, the sampling rate, the number of time points, EEG channels, targets, and harmonic frequencies considered, respectively. Also, multichannel EEG data $\mathbf{X} \in \mathbb{R}^{M \times N_t}$ and the reference signals $\mathbf{Y}_k \in \mathbb{R}^{2N_h \times N_t}$ (which are sinusoidal signals) are defined as follows:

$$\mathbf{Y}_k = [\mathbf{y}(t_1)\,\mathbf{y}(t_2)\ldots\mathbf{y}(t_{N_t})], \tag{1}$$

$$\mathbf{y}(t) = \begin{pmatrix} \sin(2\pi f_K t) \\ \cos(2\pi f_K t) \\ \vdots \\ \sin(2\pi N_h f_K t) \\ \cos(2\pi N_h f_K t) \end{pmatrix}, \quad t = \frac{1}{F_s}, \frac{2}{F_s}, \ldots, \frac{N_t}{F_s}$$

CCA finds the weight vectors $\mathbf{w}_x$ and $\mathbf{w}_y$ so that the correlation between two canonical variables $x = \mathbf{X}^\mathbf{T} \mathbf{w}_x$ and $y = \mathbf{Y}_k^T \mathbf{w}_y$ (which are linear combinations of $\mathbf{X}$ and $\mathbf{Y}_k$ respectively) is maximized by solving the following optimization problem [16]:

$$\rho_k = \max_{\mathbf{w}_x, \mathbf{w}_y} \rho(x, y) = \frac{E[x^T y]}{\sqrt{E[x^T x] E[y^T y]}} = \frac{E[\mathbf{w}_x^T \mathbf{X} \mathbf{Y}_k^T \mathbf{w}_y]}{\sqrt{E[\mathbf{w}_x^T \mathbf{X} \mathbf{X}^\mathbf{T} \mathbf{w}_x] E[\mathbf{w}_y^T \mathbf{Y}_k \mathbf{Y}_k^T \mathbf{w}_y]}} \tag{2}$$

where $\rho(x, y)$ indicates the Pearson's correlation coefficient between $x$ and $y$. Maximum of $\rho$ with respect to $\mathbf{w}_x$ and $\mathbf{w}_y$ is the maximum canonical correlation ($\rho_k$). To recognize the frequency of SSVEP, $\rho_k$ is calculated for all targets ($k = 1, 2, \ldots, K$) and the target with the maximal $\rho_k$ is selected:

$$k^* = \arg\max_k \rho_k, \quad k = 1, 2, \ldots, K \tag{3}$$

### 2.3.2. Extended CCA-Based Method

The standard CCA method is an unsupervised method, meaning that it does not use any calibration data for the target identification. This method has been originally developed for frequency detection. Since phase detection requires training data, CCA cannot be used to distinguish different phases [7]. Incorporating training data in target identification methods can capture the temporal features of SSVEP response more effectively and enhance the performance of CCA-based approaches [12, 22]. Extended CCA which combines standard CCA and individual training-based methods has been proposed in several studies [5, 7, 12, 30] and its superiority over other CCA-based training methods has been shown in [22]. In this method, individual SSVEP template signals $\hat{\mathbf{X}}_k$ are derived by averaging multiple training trials related to the $k$-th target. Then, projections of a test data $\mathbf{X}$ and an individual template $\hat{\mathbf{X}}_k$ are computed using CCA-based spatial filters, and finally, the correlation coefficients between some pairs of these projections are used as features to identify the target. Specifically, in extended CCA, four additional features are used:

$$\mathbf{r}_k = \begin{pmatrix} \mathbf{r}_k(1) \\ \mathbf{r}_k(2) \\ \mathbf{r}_k(3) \\ \mathbf{r}_k(4) \\ \mathbf{r}_k(5) \end{pmatrix} = \begin{pmatrix} \rho\left(\mathbf{X}^T \mathbf{w}_x(\mathbf{XY}_k), \mathbf{Y}_k^T \mathbf{w}_{\mathbf{Y}_k}(\mathbf{XY}_k)\right) \\ \rho\left(\mathbf{X}^T \mathbf{w}_x(\mathbf{X}\hat{\mathbf{X}}_k), \hat{\mathbf{X}}_k^T \mathbf{w}_x(\mathbf{X}\hat{\mathbf{X}}_k)\right) \\ \rho\left(\mathbf{X}^T \mathbf{w}_x(\mathbf{XY}_k), \hat{\mathbf{X}}_k^T \mathbf{w}_x(\mathbf{XY}_k)\right) \\ \rho\left(\mathbf{X}^T \mathbf{w}_{\hat{\mathbf{X}}_k}(\hat{\mathbf{X}}_k \mathbf{Y}_k), \hat{\mathbf{X}}_k^T \mathbf{w}_{\hat{\mathbf{X}}_k}(\hat{\mathbf{X}}_k \mathbf{Y}_k)\right) \\ \rho\left(\hat{\mathbf{X}}_k^T \mathbf{w}_x(\mathbf{X}\hat{\mathbf{X}}_k), \hat{\mathbf{X}}_k^T \mathbf{w}_{\hat{\mathbf{X}}_k}(\mathbf{X}\hat{\mathbf{X}}_k)\right) \end{pmatrix} \quad (4)$$

Here, $\mathbf{w}_A(AB)$ generally indicates the spatial filter derived from CCA between two multidimensional variables A and B and related to variable A. Then, linear combination of these five correlation values are used as the final feature for target identification:

$$\rho_k = \sum_{i=1}^{5} \mathbf{r}_k(i) \quad , \quad k = 1, 2, ..., K \quad (5)$$

Equation (5) also captures the discriminative information from negative correlation coefficients (all except $\mathbf{r}_k(1)$ can be negative). Although the original method uses the sum of squares of coefficients along with their signs, in this study, equation (5) is used due to its superiority in performance. Finally, the stimulus target is identified by equation (3).

### 2.3.3. TRCA-Based Method

TRCA was originally proposed in functional neuroimaging [21] and then used in SSVEP-based BCIs to obtain optimized spatial filters to improve SNR of SSVEP response [20]. The method recovers the task-related components (here SSVEP) using a linear, weighted sum of the observed signals (here, multichannel EEG signals):

$$y(t) = \sum_{j=1}^{M} w_j x_j(t) = \mathbf{w}^T \mathbf{x}(t) \quad (6)$$

where $j$ is the index of channels, $y(t) \in \mathbb{R}$ is the recovered signal, $\mathbf{x}(t) \in \mathbb{R}^M$ is the multichannel EEG signal and $\mathbf{w} \in \mathbb{R}^M$ is the optimized spatial filter derived from TRCA method. This problem can be formulated by maximizing inter-trial covariance [21]. Let $\mathbf{x}^{(h)}(t)$, $y^{(h)}(t)$, and $H$ denote $h$-th trial of $\mathbf{x}(t)$, $h$-th trial of $y(t)$, and the number of training trials, respectively. The covariance between the $h_1$-th and $h_2$-th trials of $y(t)$ is defined by:

$$C_{h_1 h_2} = \mathrm{Cov}\left(y^{(h_1)}(t), y^{(h_2)}(t)\right) = \sum_{j_1, j_2 = 1}^{M} w_{j_1} w_{j_2} \mathrm{Cov}\left(x_{j_1}^{(h_1)}(t), x_{j_2}^{(h_2)}(t)\right) \tag{7}$$

Then, the sum over all possible combinations of inter-trial covariance is considered as the objective function:

$$\sum_{\substack{h_1, h_2=1 \\ h_1 \neq h_2}}^{H} C_{h_1 h_2} = \sum_{\substack{h_1, h_2=1 \\ h_1 \neq h_2}}^{H} \sum_{j_1, j_2=1}^{M} w_{j_1} w_{j_2} \mathrm{Cov}\left(x_{j_1}^{(h_1)}(t), x_{j_2}^{(h_2)}(t)\right) = \mathbf{w}^T \mathbf{S} \mathbf{w} \tag{8}$$

To limit the weight vector in equation (8), the variance of y is normalized to one:

$$\mathrm{var}(y(t)) = \sum_{j_1, j_2=1}^{M} w_{j_1} w_{j_2} \mathrm{Cov}\left(x_{j_1}(t), x_{j_2}(t)\right) = \mathbf{w}^T \mathbf{Q} \mathbf{w} = 1 \tag{9}$$

The constrained optimization problem then becomes a Rayleigh quotient maximization:

$$\hat{\mathbf{w}} = \arg\max_{\mathbf{w}} \frac{\mathbf{w}^T \mathbf{S} \mathbf{w}}{\mathbf{w}^T \mathbf{Q} \mathbf{w}} \tag{10}$$

The optimal weight vector $\hat{\mathbf{w}}$ is equivalent to the eigenvector corresponding to the largest eigenvalue of the matrix $\mathbf{Q}^{-1}\mathbf{S}$. Then, the following correlation coefficient is computed:

$$\rho_k = \rho\left(\mathbf{X}^T \mathbf{w}_k, \hat{\mathbf{X}}_k^T \mathbf{w}_k\right) \tag{11}$$

where similar to section 2.3.2, $\mathbf{X}$ and $\hat{\mathbf{X}}_k$ are the single-trial test data and SSVEP template signal computed by averaging across trials of the $k$-th target, respectively. Also, $\mathbf{w}_k$ is the spatial filter derived from applying TRCA algorithm on the training data for the $k$-th visual stimulus. In the end, the target can be recognized by the rule provided in equation (3).

An ensemble TRCA was also proposed in [20] in which spatial filters derived for different visual stimulus were integrated to construct an ensemble of spatial filters $\mathbf{W} \in \mathbb{R}^{M \times K}$:

$$\mathbf{W} = [\mathbf{w}_1 \ \mathbf{w}_2 \ ... \ \mathbf{w}_K] \tag{12}$$

Since the mixing coefficients from the SSVEP source to the scalp recordings are approximately similar for the utilized frequency range, the $K$ different spatial filters can be considered similar, and this is the reason why ensemble TRCA is effective [20]. Equation (11) is modified as follows:

$$\rho_k = \psi\left(\mathbf{X}^{\mathbf{T}}\mathbf{W}, \hat{\mathbf{X}}_k^T \mathbf{W}\right) \tag{13}$$

where $\psi(A, B)$ indicates the two-dimensional correlation coefficient between $A$ and $B$. Again, equation (3) is used for target identification.

## 2.4. Proposed Method

Extended CCA has shortcomings. First, there are numerous ways to project training or test data on the CCA-based spatial filters and compute the correlation between each pair of these projections. Extended CCA uses only five of such correlation coefficients (equation (4)) and it is not clear how these five features are selected and the others ignored. Second, there is no ensemble extension for this method (or for any other CCA-based methods). Therefore, these methods cannot compete with ensemble TRCA which has the best performance among the current methods. To mitigate these limitations, in this study, a new method is proposed in which the best CCA-based features are selected. Moreover, to enhance the performance of the method, its ensemble version is also proposed. The structures of the proposed algorithms are illustrated in figure 1 and their details are presented below.

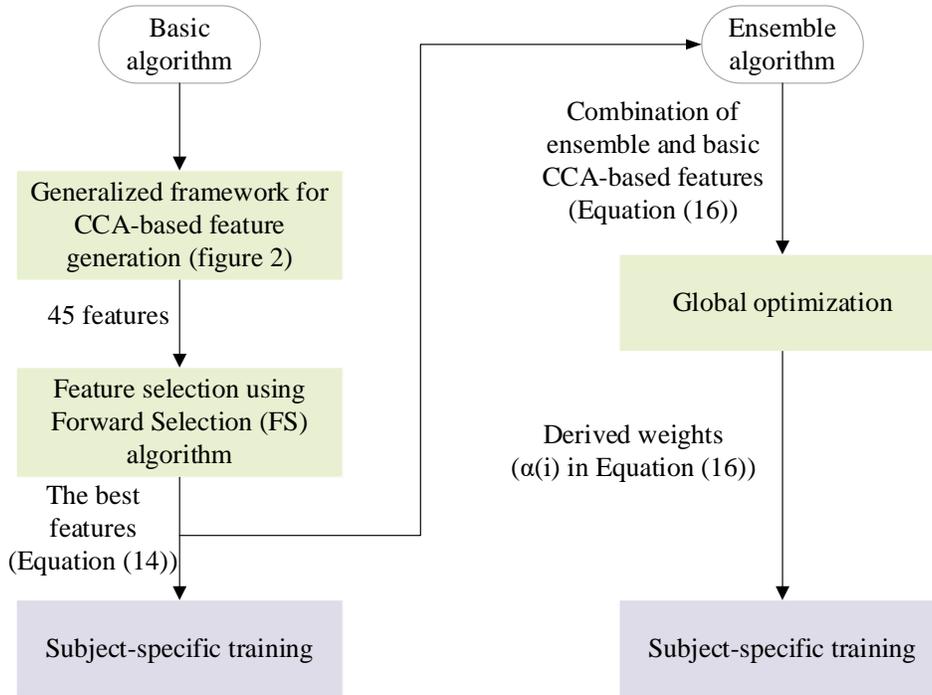

Figure 1: Structure of the proposed method and its ensemble version (green background represents subject-independent training and purple background represents subject-specific training).

### 2.4.1. Basic Algorithm

In the first step, all possible canonical variables (CVs) derived from CCA-based spatial filters are constructed. In CCA-based methods, there are three types of data including: 1- test data $\mathbf{X}$; 2- template signal $\hat{\mathbf{X}}_k$ derived from averaging across training blocks of the $k$-th target; 3- sinusoidal signals $\mathbf{Y}_k$. By computing CCA between each pair of these three types, six spatial filters are generated: 1- $\mathbf{W_X}(\mathbf{X}\hat{\mathbf{X}}_k)$; 2- $\mathbf{W}_{\hat{\mathbf{X}}_k}(\mathbf{X}\hat{\mathbf{X}}_k)$; 3- $\mathbf{W_X}(\mathbf{XY}_k)$; 4- $\mathbf{W}_{\hat{\mathbf{X}}_k}(\hat{\mathbf{X}}_k\mathbf{Y}_k)$; 5- $\mathbf{W}_{\mathbf{Y}_k}(\mathbf{XY}_k)$; and 6- $\mathbf{W}_{\mathbf{Y}_k}(\hat{\mathbf{X}}_k\mathbf{Y}_k)$. Projections of $\mathbf{X}$ and $\hat{\mathbf{X}}_k$ on the first four spatial filters and $\mathbf{Y}_k$ on the 5th and 6th spatial filters generate a total of 10 CVs. These CVs are listed in table 1.

In the second step, the best correlation features derived from the correlation between each pair of the CVs are found. Since there are 10 CVs, 45 correlation features can be computed ($\binom{10}{2} = 45$).

Figure 2 shows the block diagram of the proposed method for generating the 45 correlation features. Most of these features can be used for target identification. The correlation coefficients between projections of $\hat{\mathbf{X}}_k$ and projections of $\mathbf{Y}_k$ (including 8 features) have no capability of detecting SSVEPs even if test data are used to construct spatial filters. Also, the correlation between CV9 and CV10 is not useful. Therefore, any combination of the remaining 36 features can be selected for subject-specific training. There are a variety of feature selection algorithms in the literature [31, 32]. In this paper, a simple feature selection algorithm called forward selection (FS) [32] is used to find the best set of correlation features. In this algorithm, the feature which maximizes the average classification accuracy among the 36 features is selected. The classification measure is the same as the one presented in equation (3). Then, the second feature is selected such that the features selected in the previous and present steps lead to best performance. Similar to equation (5), the summation of features is used to combine features for classification. The process of adding features continues until there is no improvement in average classification accuracy. Finally, the feature set in the last step is recognized as the best feature set.

The subject independent training is employed to create 45 features and apply FS algorithm on them (see section 2.4.3). Therefore, after applying FS algorithm on seven folds described in 2.4.3, seven feature sets that contain best features for each fold are obtained. The interesting point is that in all these feature sets, the maximum performance is provided by six features which are the same across different folds. However, the order in which these features are selected is not same. Further information regarding features selected in each fold can be found in the Appendix A. These six best features are as follows:

$$\mathbf{r}_k = \begin{pmatrix} \mathbf{r}_k(1) \\ \mathbf{r}_k(2) \\ \mathbf{r}_k(3) \\ \mathbf{r}_k(4) \\ \mathbf{r}_k(5) \\ \mathbf{r}_k(6) \end{pmatrix} = \begin{pmatrix} \rho\left(\mathbf{X}^T\mathbf{W_X}(\mathbf{XY}_k), \mathbf{Y}_k^T\mathbf{W}_{\mathbf{Y}_k}(\mathbf{XY}_k)\right) \\ \rho\left(\mathbf{X}^T\mathbf{W_X}(\mathbf{X}\hat{\mathbf{X}}_k), \hat{\mathbf{X}}_k^T\mathbf{W}_{\hat{\mathbf{X}}_k}(\mathbf{X}\hat{\mathbf{X}}_k)\right) \\ \rho\left(\mathbf{X}^T\mathbf{W_X}(\mathbf{X}\hat{\mathbf{X}}_k), \hat{\mathbf{X}}_k^T\mathbf{W_X}(\mathbf{X}\hat{\mathbf{X}}_k)\right) \\ \rho\left(\mathbf{X}^T\mathbf{W_X}(\mathbf{XY}_k), \hat{\mathbf{X}}_k^T\mathbf{W_X}(\mathbf{XY}_k)\right) \\ \rho\left(\mathbf{X}^T\mathbf{W}_{\hat{\mathbf{X}}_k}(\hat{\mathbf{X}}_k\mathbf{Y}_k), \hat{\mathbf{X}}_k^T\mathbf{W}_{\hat{\mathbf{X}}_k}(\hat{\mathbf{X}}_k\mathbf{Y}_k)\right) \\ \rho\left(\mathbf{X}^T\mathbf{W}_{\hat{\mathbf{X}}_k}(\hat{\mathbf{X}}_k\mathbf{Y}_k), \mathbf{Y}_k^T\mathbf{W}_{\mathbf{Y}_k}(\hat{\mathbf{X}}_k\mathbf{Y}_k)\right) \end{pmatrix} \qquad (14)$$

The coefficients $\mathbf{r}_k(1)$, $\mathbf{r}_k(3)$, $\mathbf{r}_k(4)$, and $\mathbf{r}_k(5)$ are present in both extended CCA and the proposed method while the coefficients $\mathbf{r}_k(2)$ and $\mathbf{r}_k(6)$ are exclusively present in our method. These coefficients are used for subject-specific training in the basic algorithm. Similar to equation (5), the following relation is used to build the final feature for classification (equation (3)):

$$\rho_k = \sum_{i=1}^{6} r_k(i) \quad , \quad k = 1, 2, \ldots, K \tag{15}$$

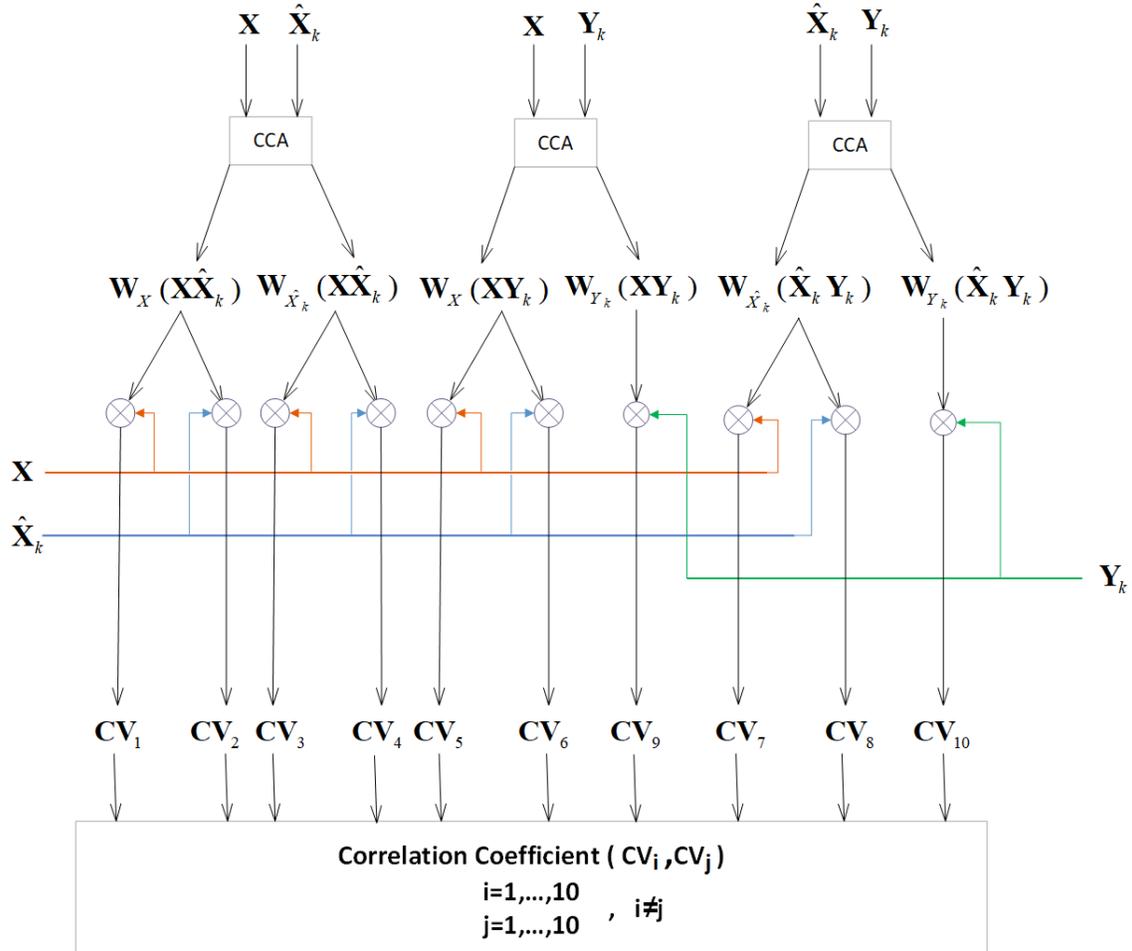

Figure 2: Block diagram for generating all possible CCA-based correlation features.

Table 1: Mathematical description of the 10 CVs depicted in figure 2.

| Canonical Variable | Formula | Canonical Variable | Formula |
|---|---|---|---|
| $CV_1$ | $\mathbf{X^T W_X (X\hat{X}}_k)$ | $CV_6$ | $\mathbf{\hat{X}}_k^T \mathbf{W_X (XY}_k)$ |
| $CV_2$ | $\mathbf{\hat{X}}_k^T \mathbf{W_X (X\hat{X}}_k)$ | $CV_7$ | $\mathbf{X^T W_{\hat{X}_k} (\hat{X}}_k \mathbf{Y}_k)$ |
| $CV_3$ | $\mathbf{X^T W_{\hat{X}_k} (X\hat{X}}_k)$ | $CV_8$ | $\mathbf{\hat{X}}_k^T \mathbf{W_{\hat{X}_k} (\hat{X}}_k \mathbf{Y}_k)$ |
| $CV_4$ | $\mathbf{\hat{X}}_k^T \mathbf{W_{\hat{X}_k} (X\hat{X}}_k)$ | $CV_9$ | $\mathbf{Y}_k^T \mathbf{W_{Y_k} (XY}_k)$ |
| $CV_5$ | $\mathbf{X^T W_X (XY}_k)$ | $CV_{10}$ | $\mathbf{Y}_k^T \mathbf{W_{Y_k} (\hat{X}}_k \mathbf{Y}_k)$ |

### 2.4.2. Ensemble Algorithm

Ensemble TRCA showed that an integration of spatial filters derived from calibration data of different classes enhanced performance of SSVEP BCIs [20]. In fact, using both between and within class information in pattern classification methods can boost classifier performance [32]. According to equation (13), to exploit an ensemble of spatial filters for a correlation-based feature between two sets, two conditions must be satisfied. First, these two sets should be projected on the same group of spatial filters. Second, the group must contain spatial filters of all classes. By evaluating these two conditions for the six features in equation (14), only $\mathbf{r}_k(3)$, $\mathbf{r}_k(4)$, and $\mathbf{r}_k(5)$ satisfy the first condition and only $\mathbf{r}_k(5)$ satisfies the second condition. Consequently, the six features $\mathbf{r}_k$ in equation (14) can be converted to the six features $\hat{\mathbf{r}}_k$ in which all features are same as $\mathbf{r}_k$ except $\hat{\mathbf{r}}_k(5)$. This feature is constructed using the two-dimensional correlation between two projections on the ensemble of spatial filters derived from CCA between template signals $\hat{\mathbf{X}}_k$ and sinusoidal signals $\mathbf{Y}_k$. Since $\hat{\mathbf{r}}_k(5)$ is the best discriminative feature compared with the other coefficients, the uniform combination of the six coefficients, similar to equation (15), does not seem to be the best possible solution. To take feature differences into account, a linear weighted sum of coefficients $\hat{\mathbf{r}}_k(i)$ is proposed:

$$\rho_k = \sum_{i=1}^{6} \alpha(i).\hat{\mathbf{r}}_k(i) \quad , \quad k = 1, 2, ..., K \tag{16}$$

The mixing weights $\alpha(i)$ are estimated using subject independent data (see section 2.4.3). The objective function is maximization of average classification accuracy, computed based on equations (3) and (16). Since this combination is a complex nonlinear function in terms of $\alpha(i)$, gradient-based optimization methods cannot be applied. Considering the limited parameter space of the problem, metaheuristic optimization methods including Genetic (GA) or particle swarm optimization (PSO) can be used. We use a GA algorithm to estimate $\alpha(i)$ coefficients such that the objective function is maximized. GA is implemented using the ga function in MATLAB. For the sake of simplicity and limiting the search space, the coefficients are confined inside the [0 1] interval. Based on this procedure, it is trivial that the estimation process will provide the largest weight ($\alpha(5)$) for $\hat{\mathbf{r}}_k(5)$. Finally, it should be noted that the estimated weights $\alpha(i)$ are obviously different for each fold.

### 2.4.3. Cross-Validation

As mentioned before, both subject-independent and subject-specific training are used in the proposed method. Cross-validation is performed on the subjects and the six blocks of a specific subject for the first and second training techniques, respectively. Further information about cross-validation techniques is presented next.

**Subject-independent training:** The parts related to this training technique are shown in green in figure 1. In this approach, the K-fold (K=7) approach is used and the data of 30 subjects is utilized to obtain the best hyperparameters for the remaining 5 subjects. Then, the obtained hyperparameters are employed to create the subject-specific models. Specifically, in the basic algorithm, for each fold, 45 CCA-based features are constructed for 30 subjects and then, the features that maximize average recognition accuracy for the mentioned subjects are selected (section 2.4.1). Finally, the subject-specific models are created for 5 remaining subjects using the selected features. Similarly, in the ensemble algorithm, the weights (section 2.4.2) that maximize the average accuracy for the 30 subjects of the corresponding fold are used to build the subject-specific models of the remaining subjects. Therefore, the selected features in the basic algorithm and the weights ($\alpha(i)$) in the ensemble algorithm are the hyperparameters.

**Subject-specific training:** In both basic and ensemble algorithms, subject-specific models are built using the hyperparameters derived from other subjects' data. For each subject, the leave-one-out technique is used on the six blocks. In other words, the data samples from five blocks are used as the training data to construct a reference signal for each target while the left out single block is used for validation. This procedure is repeated six times such that every block is considered as validation data once. Finally, the average recognition accuracy across these six blocks are computed. It is worthwhile to note that the classification accuracies reported in the result section only refer to this type of training.

### 2.5. Filter Bank Analysis

Higher harmonics of the SSVEP stimulus frequency contain useful information which can improve the recognition accuracy. To extract this information, filter bank analysis has been proposed as a practical solution in which a signal is decomposed to multiple frequency sub-bands [29, 34]. Filter bank analysis can reduce the detection error due to background EEG activities. X. Chen, et al. [29] applied the filter bank technique to the SSVEP-based BCIs, enhancing performance of standard CCA method significantly. This technique is applied to all methods presented here and its effect is reported. To design the filter bank, a procedure similar to [12, 29] is utilized. In this method, the EEG data is decomposed to N sub-bands using the N band-pass filters and a feature extraction algorithm is applied to each sub-band separately. The lower and upper cut-off frequencies of the *n*-th sub-band are set to n×8 Hz and 70 Hz, respectively . The zero-phase Chebyshev Type II IIR band-pass filter is used to extract every sub-band signals. The features computed from the sub-bands are combined as follows:

$$\tilde{\rho}_k = \sum_{n=1}^{N} w_{SB}(n)\rho_k^{(n)} \tag{17}$$

where $\rho_k^{(n)}$, $\tilde{\rho}_k$, and $w_{SB}(n)$ indicate the feature value for the *n*-th sub-band and *k*-th target, the final feature for classification, and the weights for the sub-band components, respectively. Based on the previous studies, when response frequency increased, SNR of SSVEP decreased [29]. Therefore, the sub-band weights are determined using the following equation:

$$w_{SB}(n) = n^{-a} + b, \quad n \in [1 \ N] \tag{18}$$

Following [12], *a* and *b* are set to 1 and 0, respectively. As mentioned before, the target is selected by equation (3) and substituting $\rho_k$ with $\tilde{\rho}_k$.

## 3. Results

Classification accuracy and ITR were used as evaluation metrics to compare performance of different methods. These two metrics were calculated with different data lengths (from 0.2 s to 1 s with a step of 0.1 s). The 0.5 s gaze shifting duration was considered to compute the simulated ITR in the offline analysis. Also, the number of harmonics in equation (1) was set to 3. Figure 3 shows the average accuracies and ITRs across subjects for three basic methods at different time windows, with and without filter bank. For the filter bank, the number of sub-bands was set to 4. In all possible cases, TRCA showed superior performance over other methods for time windows shorter than 0.3 s. In 0.3 s, one-way repeated measures analysis of variance (ANOVA) showed no significant difference between the accuracy ($F(2,68)=1.35$, $p=0.26$) and ITR ($F(2,68)=1.09$, $p=0.33$) of the three methods without filter bank. When filter bank was applied to the methods in 0.3 s, ANOVA revealed significant difference in both accuracy ($F(2,68)=17.79$, $p<0.001$) and ITR ($F(2,68)=18.45$, $p<0.001$) between the three methods. Post-hoc paired t-tests showed that there was no significant difference in accuracy ($p=0.67$) and ITR ($p=0.62$) between TRCA and the proposed method while both methods outperformed extended CCA ($p<0.001$). For time windows greater than 0.3 s, ANOVA indicated significant difference ($p<0.01$) between the three methods in all conditions. Post-hoc paired t-tests confirmed superior performance of the proposed method relative to TRCA and extended CCA ($p<0.01$). In figure 3(b), the time windows corresponding to the highest ITR were different for each method (extended CCA: 0.8 s; TRCA: 0.8 s; the proposed method: 0.7 s) while in figure 3(d), all methods reached their highest ITR in 0.7 s.

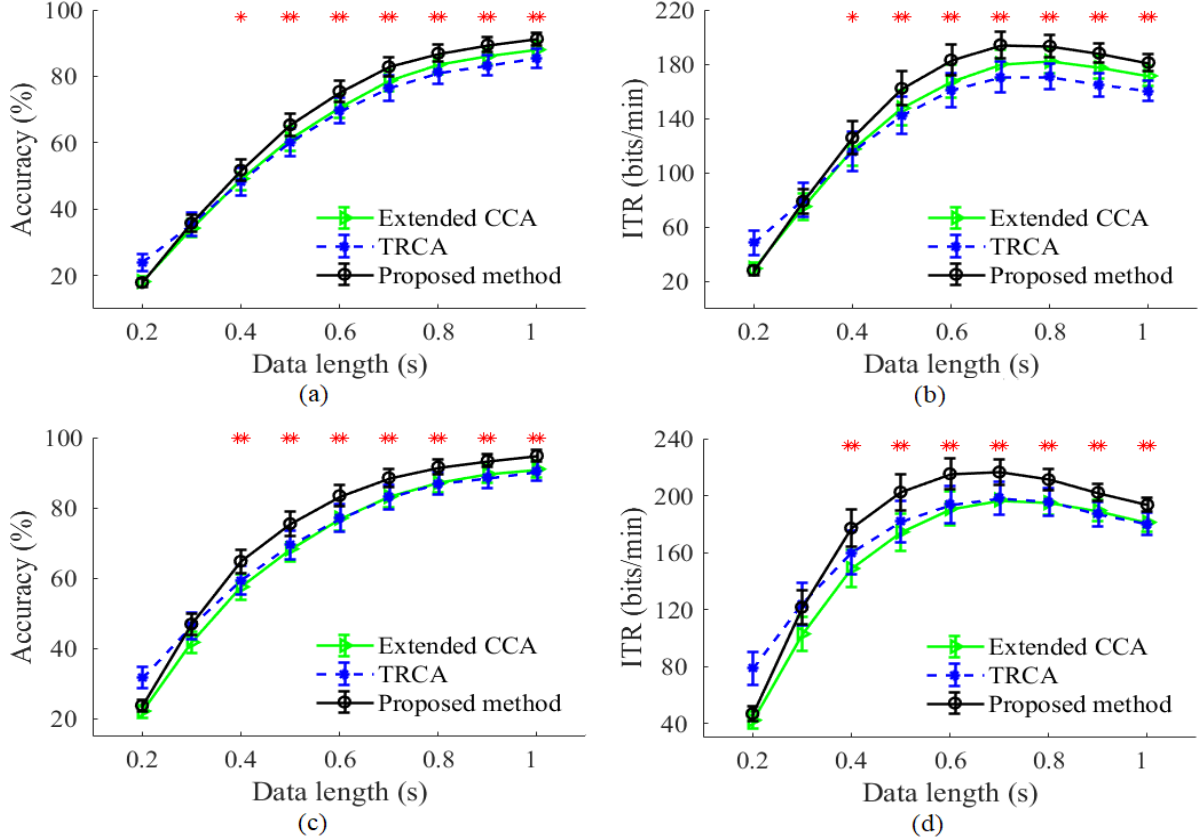

Figure 3: Average accuracies, (a) and (c), and ITRs, (b) and (d), across subjects for three basic methods at different time windows. Results in first and second rows are derived without and with filter bank, respectively. Number of sub-bands is set to 4. Asterisks represent significant difference between the three methods, using ANOVA at time windows greater than 0.3 (*p<0.01, **p<0.001). Error bars show standard errors.

The ensemble version of the proposed method is compared with ensemble TRCA in figure 4. To estimate the combination weights, i.e., ($\alpha(i)$), in equation (16) using the procedure described in section 2.4.2, time window was set to 0.5 s. Similar to the basic methods, ensemble TRCA performed better than the proposed ensemble method in all cases when data length was less than 0.3 s. For 0.3 s, paired t-tests showed no significant difference between the two methods, with and without filter bank (figure 4(a): p=0.62; figure 4(b): p=0.50; figure 4(c): p=0.12; figure 4(d): p=0.35). For data lengths greater than 0.3 s, the proposed ensemble method led to significantly (p<0.001) higher accuracy and ITR than ensemble TRCA for both cases. Both methods reached their highest ITRs at 0.6 s in figure 4(b) and 0.5 s in figure 4(d).

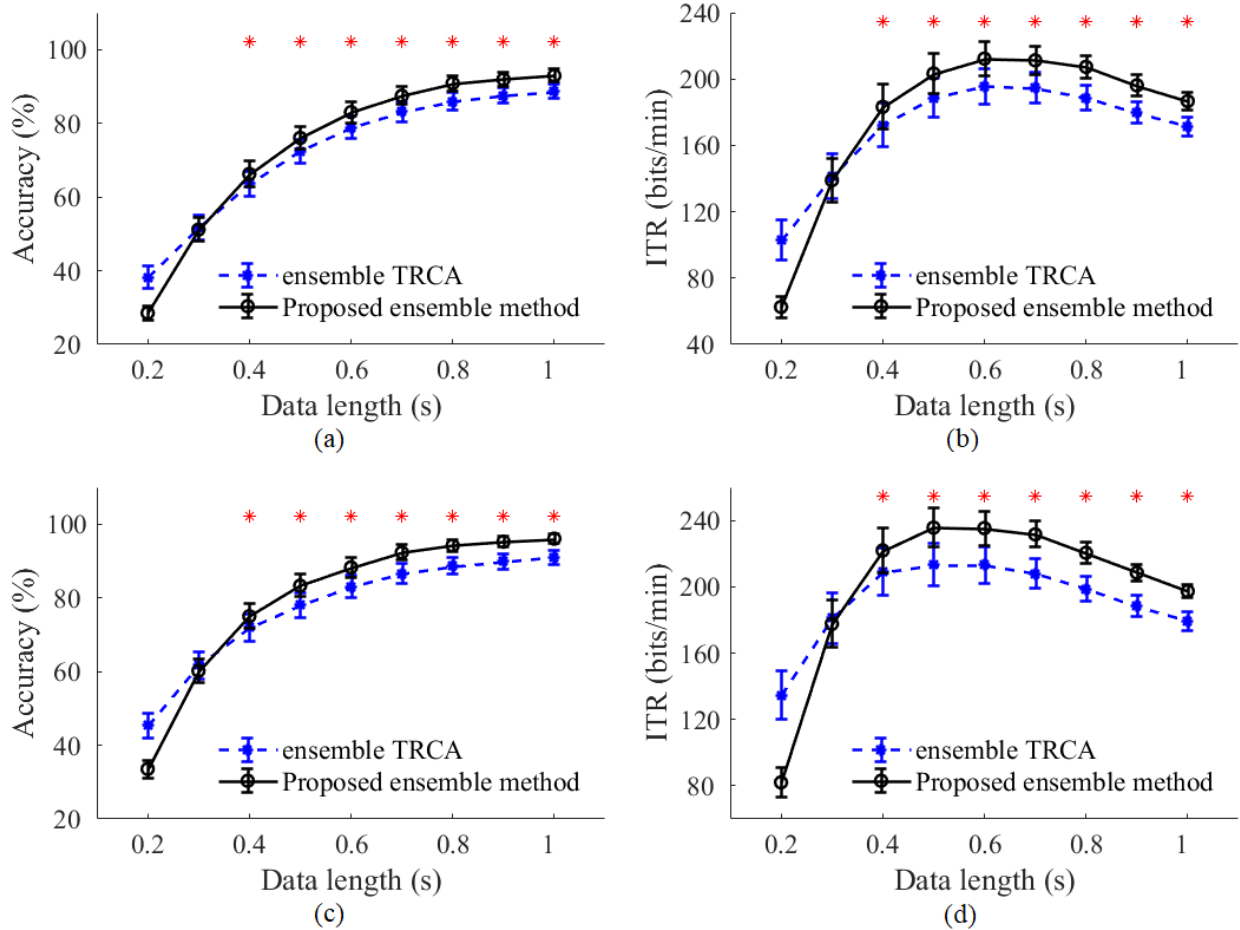

Figure 4: Average accuracies, (a) and (c), and ITRs, (b) and (d), across subjects for ensemble TRCA and ensemble version of the proposed method at different time windows. Results in first and second rows are derived without and with filter bank, respectively. Number of sub-bands is set to 4. Asterisks represent significant difference between the two methods by paired t-tests at time windows greater than 0.3 (*p<0.001). Error bars show standard errors.

Performance of different training methods depends on the number of sub-bands, electrodes, and training blocks. Therefore, the effects of varying these parameters on the classification accuracy for all cases including basic and ensemble TRCA, and basic and ensemble proposed method are investigated in figures 5 and 6. Time window was 0.5 s to perform the analysis. In figure 5, the number of training blocks and electrodes were fixed at 5 and 9 and the effect of the number of sub-bands was explored. The proposed method represents significantly (p<0.001) higher classification accuracies than TRCA in all cases. For both basic and ensemble versions of the two methods, the highest accuracy is achieved by 4 sub-bands. According to this fact, the number of sub-bands was fixed at 4 and the variations of average accuracies corresponding to different number of electrodes and training blocks were examined in figure 6. The results illustrate that for both basic and ensemble cases, the proposed method improves TRCA, especially for low number of training blocks and electrodes (p<0.001). Furthermore, TRCA needs at least two training blocks to obtain optimal spatial filters while the proposed method can deliver an acceptable performance even with a single training block (see figures 6(b) and 6(d)). This characteristic can be one of the major advantages of our method compared with TRCA. Typically, in SSVEP BCI, it is necessary to collect training data at the beginning of each session which could be time-consuming; our method reduces training time considerably.

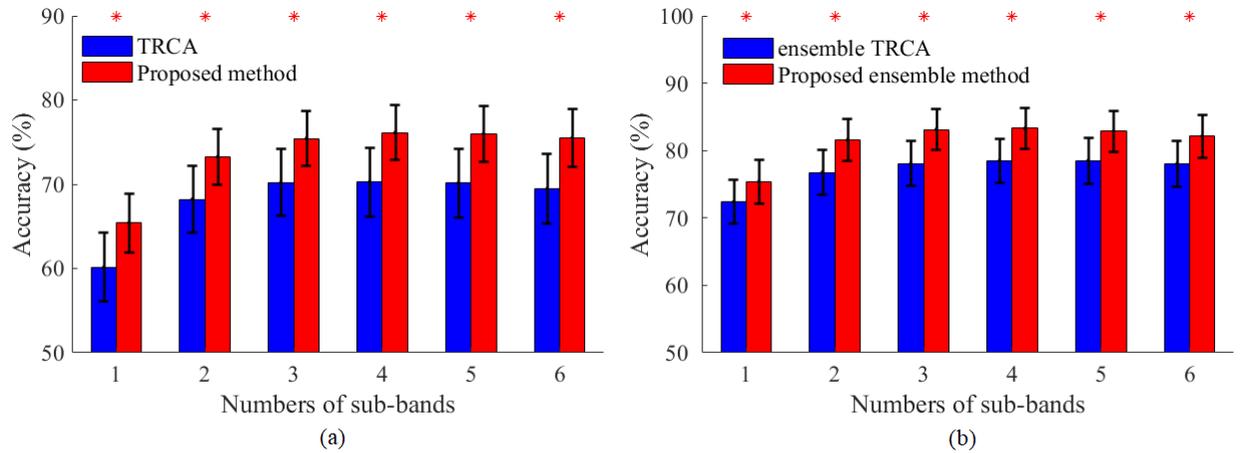

Figure 5: Average accuracies across subjects for different number of sub-bands: (a) basic TRCA and the proposed method; and (b) ensemble TRCA and the proposed ensemble method. Asterisks represent significant difference between the two methods by paired t-tests (*p<0.001). Error bars show standard errors.

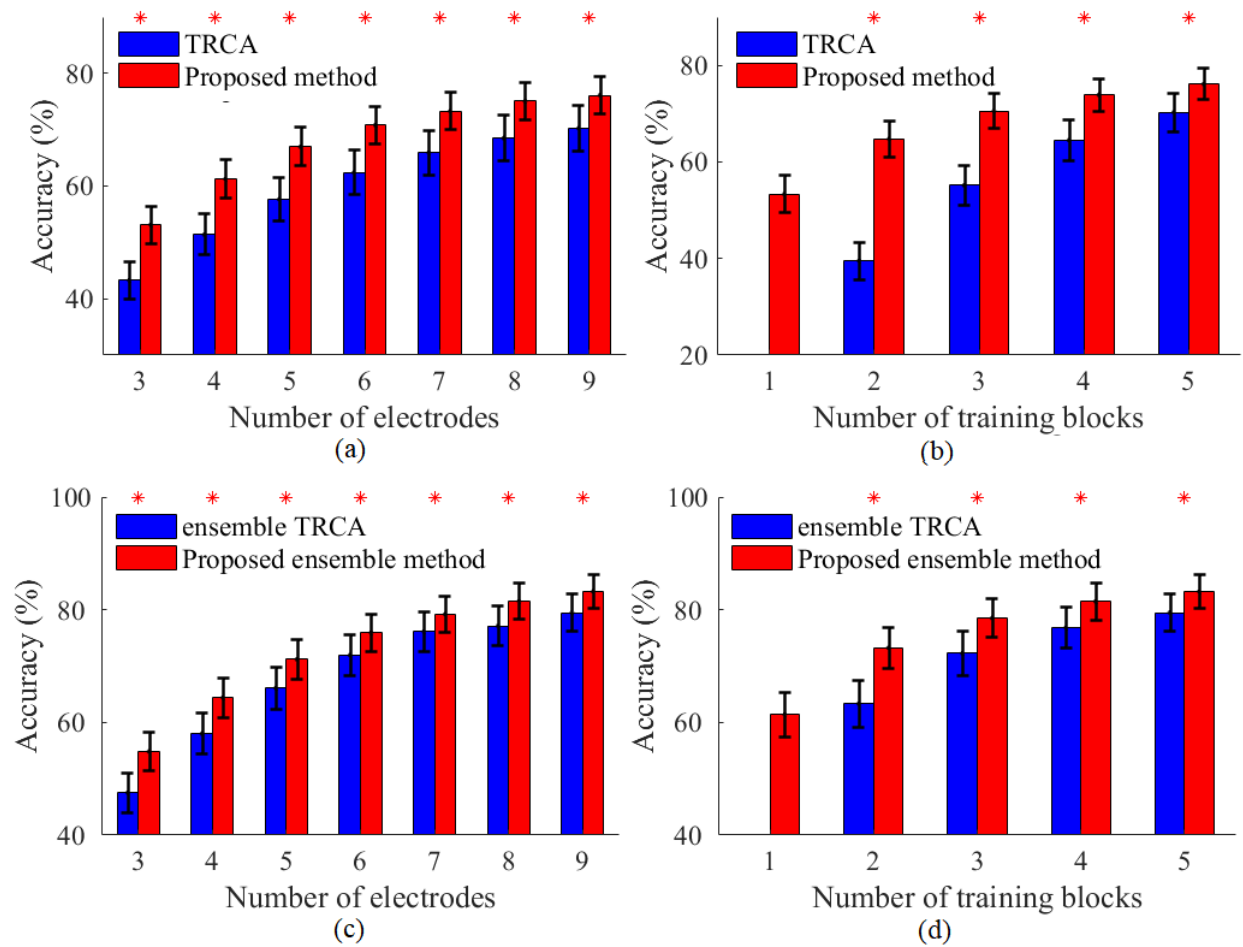

Figure 6: Average accuracies across subjects obtained by different number of electrodes, (a) and (c), and training blocks, (b) and (d). First row compares two basic methods and second row compares two ensemble methods. Asterisks represent significant difference between the two methods by paired t-tests (*p<0.001). Error bars show standard errors.

## 4. Discussion

Classification accuracy and ITR are the most important factors for practical development of SSVEP-based BCI spellers and thus must be improved as much as possible. In this study, an ensemble CCA-based training method was proposed for the first time, which improved the performance of the extended CCA and TRCA methods. The proposed method outperformed extended CCA in all conditions. Furthermore, it improved TRCA method in terms of both accuracy and ITR for data lengths greater than 0.3 s. Lower performance of our method for short time durations could be related to inaccurate estimation of the spatial filters obtained by the CCA algorithm in data with a small sample size. However, when the data length increases, on one hand, spatial filters are estimated more accurately and on the other hand, the combination of various coefficients which exploit CCA-based spatial filters improve the performance of the proposed method compared with TRCA method. In practical applications, for majority of subjects, the maximum speed (highest ITR) is reached at time windows greater than 0.3, justifying the application of the proposed method for such subjects. All in all, the only case that TRCA method is preferable to the proposed method is when the number of blocks and electrodes are large and the subject reaches his/her highest ITR in 0.3 s or less. Otherwise, the proposed method is recommended. Also, in this paper, due to the limited number of training blocks per subject, the subject-independent training technique was used to find the best CCA-based features and estimate mixing weights in equation (16). However, for a new subject, equations (14), (15), (16), and one set of weights $\alpha(i)$ are sufficient for target detection.

For further investigation of the performance of the proposed method relative to TRCA, feature values can be compared for the two methods. Since the scale of final features obtained by the two methods is different, feature vectors derived from each trial are linearly normalized into [-1, +1] and then compared. Figures 7(a) and 7(b) represent normalized feature values for a sample frequency derived from two basic and two ensemble methods, respectively. The number of sub-bands, electrodes, and training blocks were 4, 9, and 5, respectively. A short data length (0.6 s) was selected to carry out comparisons. In both figures, the feature values of the two methods decline with a similar trend in the neighborhood of the true frequency. However, as we move away from the true frequency, feature values of the proposed method become significantly (p<0.001) lower than TRCA method. Therefore, probability of false detection for our method is lower, leading to its further performance improvement relative to TRCA.

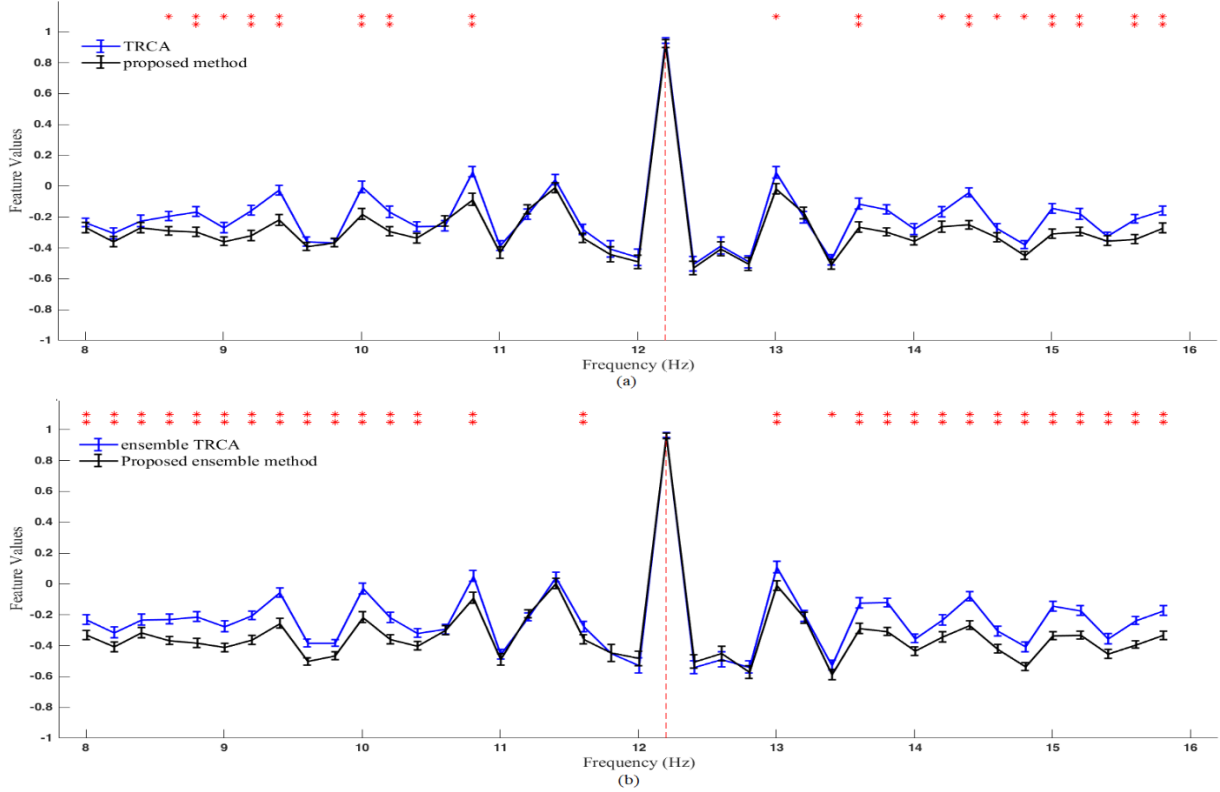

Figure 7: An example of normalized feature values, averaged across subjects and blocks, obtained by: (a) two basic methods; and (b) two ensemble methods. Red vertical line indicates true frequency. Data length is 0.6 s. Asterisks represent a significant difference between the two methods by paired t-tests (*p<0.01, **p<0.001). Error bars show standard errors.

There are several parameters in this paper which can be further optimized for each method (or subject) separately, including filter bank design, stimulus design, and electrode setting. As a representative example, consider different possible set of *n* (*n<9*) electrodes which can be selected from the nine electrodes introduced in section 2.2. For an *n*, the optimal electrode layout per method can be found by grid search, i.e., by calculating average accuracies across subjects for each layout and selecting the layout with the highest accuracy. This analysis is done on the benchmark dataset with three to six electrodes for the proposed ensemble method and ensemble TRCA. Then, the best layout per method along with the corresponding accuracies are shown in figure 8(a). This figure shows that by selecting a suitable subset of four or five electrodes, acceptable accuracies, comparable with those obtained by nine electrodes, can be achieved. It also illustrates that if we consider a local area (i.e., visual area), the best layout obtained by grid search, is almost independent of spatial filter-based target identification methods.

Another approach for optimizing electrode settings is channel selection in an unsupervised manner [35]. Maximum achievable accuracy per subject derived by grid search can be used as a reference to compare performance of channel selection algorithms in future studies. For example, figure 8(b) shows average accuracies after selecting the best electrodes per subject. This figure reveals the great potential of an effective channel selection algorithm to enhance the performance of the methods. Superior performance of the proposed method compared with TRCA is illustrated in both figures 8(a) and 8(b).

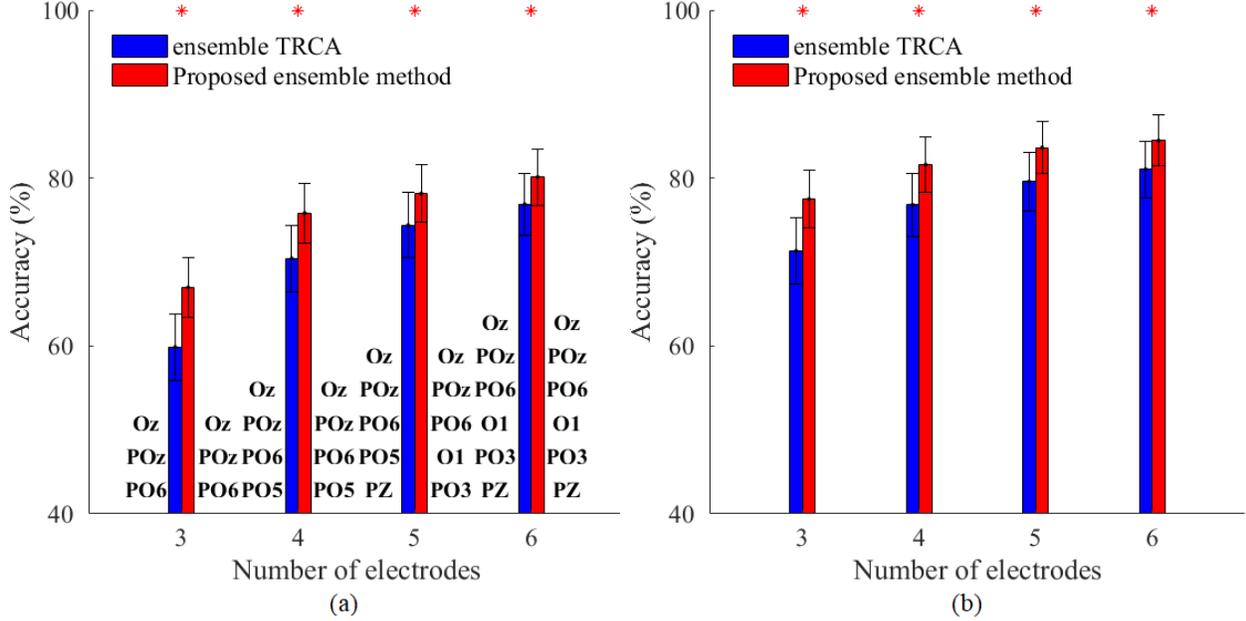

Figure 8: (a) best layout of electrodes per method, derived from grid search for all subjects and the corresponding average accuracies; and (b) potential average accuracies across subjects after selecting the best layout of electrodes per subject. In both figures, data length is 0.5 s. Asterisks represent a significant difference between the two methods by paired t-tests (*p<0.001). Error bars show standard errors.

In this study, a method was proposed which uses both subject-specific and subject-independent training techniques. Since collecting training data is time-consuming and may be exhausting for some subjects, transfer learning methods have been proposed which use training data of other subjects [24] or different sessions of the same subject [36]. Furthermore, using the benchmark dataset containing a large number of subjects [25], various training-free algorithms can be devised and evaluated in future studies to improve effectiveness of such methods. As the optimal data length for each trial can be different, an adaptive selection of time window length using a dynamic stopping criteria can be a solution for BCI users [37-38]. Besides, the combination of SSVEP and other modalities, e.g., eye-tracking systems [39] can improve the performance compared with using two single-modality methods. However, efficiency of hybrid methods over single-modality methods needs to be investigated.

## 5. Conclusion

This study proposed a framework to improve traditional CCA-based training methods by finding the best hyperparameters for each subject using other subjects' training data. These hyperparameters were used to construct basic and ensemble versions of the proposed method. The offline analysis based on a benchmark dataset was performed and the proposed method was compared with extended CCA and TRCA methods. Our method showed significantly higher performance than extended CCA in all conditions and TRCA in time windows greater than 0.3 s. All three methods can be implemented in online BCI applications to realize a high-speed SSVEP-based speller.

### Acknowledgment

The authors would like to thank the authors of [25] who provided the benchmark dataset freely.

## Appendix A. Feature Selection Results

As described in section 2.4.1, a forward selection (FS) approach was used to find the best correlation coefficients derived from CCA-based spatial filters among 36 correlation features (figure 2). This procedure was repeated for each fold using a subject-independent training schema. Feature selection was done with 0.5 s data length. Surprisingly, for each fold, the same set of features was found, and the maximum accuracy was obtained using the best six features. The results of each fold for these features are shown in table A1. CVs are the canonical variables defined in table 1, and the numbers inside the parentheses show classification accuracy for each stage of the selection process. Note that for computing the accuracy of the $k$-th fold ($k$-th row), the subjects of that fold were left out, and the classification accuracy was computed using the remaining 30 subjects.

As can be seen, although the order of selection is different for each fold, the same set of features is obtained. Cells with a blue background represent features introduced in this paper and cells with white background represent features used previously in an extended CCA method [12]. The correlation coefficient $\rho(CV_7, CV_{10})$ obtained by the proposed method is the best feature in four folds. The logic behind selection of these features is clearly described in the proposed method

Table A1: Selection of best six features for seven folds. The numbers (inside the parentheses) in the $n$-th column indicate classification accuracy with the best $n$ features.

|  | 1st coeff* | 2nd coeff | 3rd coeff | 4th coeff | 5th coeff | 6th coeff |
|---|---|---|---|---|---|---|
| fold 1 | $\rho(CV_7, CV_8)$ (51.8 %) | $\rho(CV_5, CV_6)$ (57.8 %) | $\rho(CV_1, CV_2)$ (58.9 %) | $\rho(CV_7, CV_{10})$ (60.8 %) | $\rho(CV_1, CV_4)$ (62.4 %) | $\rho(CV_5, CV_9)$ (63 %) |
| fold 2 | $\rho(CV_7, CV_{10})$ (55.1 %) | $\rho(CV_1, CV_2)$ (62.1 %) | $\rho(CV_5, CV_6)$ (63.8 %) | $\rho(CV_5, CV_9)$ (64.5 %) | $\rho(CV_7, CV_8)$ (66.7 %) | $\rho(CV_1, CV_4)$ (67 %) |
| fold 3 | $\rho(CV_7, CV_8)$ (54.2 %) | $\rho(CV_1, CV_2)$ (61 %) | $\rho(CV_5, CV_6)$ (62.5 %) | $\rho(CV_5, CV_9)$ (64.2 %) | $\rho(CV_7, CV_{10})$ (65.1 %) | $\rho(CV_1, CV_4)$ (65.8 %) |
| fold 4 | $\rho(CV_7, CV_8)$ (56.2 %) | $\rho(CV_1, CV_2)$ (62.1 %) | $\rho(CV_5, CV_6)$ (64.3 %) | $\rho(CV_5, CV_9)$ (66.1 %) | $\rho(CV_7, CV_{10})$ (67 %) | $\rho(CV_1, CV_4)$ (67.8 %) |
| fold 5 | $\rho(CV_7, CV_{10})$ (52.7 %) | $\rho(CV_1, CV_2)$ (60 %) | $\rho(CV_5, CV_6)$ (62.2 %) | $\rho(CV_5, CV_9)$ (63.3 %) | $\rho(CV_7, CV_8)$ (64 %) | $\rho(CV_1, CV_4)$ (64.6 %) |
| fold 6 | $\rho(CV_7, CV_{10})$ (53 %) | $\rho(CV_1, CV_2)$ (59.1 %) | $\rho(CV_5, CV_6)$ (62 %) | $\rho(CV_5, CV_9)$ (63.2 %) | $\rho(CV_7, CV_8)$ (64 %) | $\rho(CV_1, CV_4)$ (64.7 %) |
| fold 7 | $\rho(CV_7, CV_{10})$ (53.4 %) | $\rho(CV_1, CV_2)$ (59.5 %) | $\rho(CV_5, CV_6)$ (61.8 %) | $\rho(CV_5, CV_9)$ (63.8 %) | $\rho(CV_7, CV_8)$ (64.6 %) | $\rho(CV_1, CV_4)$ (65.2 %) |

*coeff= coefficient